\documentclass[12pt,letterpaper]{article}
\usepackage{amsmath,amssymb,pgf,pgfarrows,pgfnodes,float,appendix, hyperref}
\usepackage{graphicx}
\usepackage{subfigure}
\usepackage[margin=0.9in]{geometry}

\title{{\rm\footnotesize \qquad \qquad \qquad \qquad \qquad \ \qquad \qquad \qquad \ \ \ \ \ \                  RUNHETC-2016-15, UTTG-18-16 }\vskip.5in    Soft Gravitons and the Flat Space Limit of Anti-deSitter Space}
\author{Tom Banks\\
Department of Physics and NHETC\\
Rutgers University, Piscataway, NJ 08854\\
E-mail: \href{mailto:tibanks@ucsc.edu}{tibanks@ucsc.edu}
\\
\\
Willy Fischler\\
Department of Physics and Texas Cosmology Center\\
University of Texas, Austin, TX 78712\\
E-mail: \href{mailto:fischler@physics.utexas.edu}{fischler@physics.utexas.edu}}
\date{}
\begin{document}
\maketitle

\begin{abstract}
We argue that flat space amplitudes for the process $ 2 \rightarrow n$ gravitons at center of mass energies $\sqrt{s}$ much less than the Planck scale, will coincide approximately with amplitudes calculated from correlators of a boundary CFT for AdS space of radius $R\gg L_P$, only when $n < R/L_P$ .  For larger values of $n$ in AdS space, insisting that all the incoming energy enters ``the arena"\cite{polchsuss} , implies the production of black holes, whereas there is no black hole production in the flat space amplitudes. We also argue, from unitarity, that flat space amplitudes for all $n$ are necessary to reconstruct the complicated singularity at zero momentum in the $2 \rightarrow 2$ amplitude, which can therefore not be reproduced as the limit of an AdS calculation.  Applying similar reasoning to amplitudes for real black hole production in flat space, we argue that unitarity of the flat space S-matrix cannot be assessed or inferred from properties of CFT correlators.
  
\end{abstract}

\section{Introduction}

The study of the flat space limit of correlation functions in AdS/CFT was initiated by the work of Polchinski and Susskind\cite{polchsuss} and continued in a host of other papers.  Most of those papers agree with the contention, that the limit of Witten diagrams for CFT correlators smeared with appropriate test functions, for operators carrying vanishing angular momentum on the compact space,\footnote{All firmly established large radius examples of AdS/CFT are of the form $AdS \times K$, where $K$ is a compact manifold whose radius of curvature is of order the AdS radius.}
converge to flat space S-matrix elements between states in Fock space. Giddings\cite{giddings} has presented arguments that this consensus is incorrect.  The present authors understand that the individual lines of Feynman-Witten diagrams for Mellin transformed amplitudes converge to the required flat space limits, which suggests that if the issues pointed out by Giddings are serious, they have to do with loop integrations, but we know of no attempts to actually demonstrate the problems in an explicit calculation.
The issues discussed in this paper do not have to do with any individual Feynman-Witten diagram, but rather with the properties of such diagrams when the number of legs go to infinity.

Recently, there has been renewed attention paid to the role of ``zero momentum" gauge bosons and gravitons in scattering theory in Minkowski space\cite{tbas}.  In four dimensions, these are related to the notorious infrared (IR) divergences in scattering amplitudes for Fock space states, and the problem is ``resolved" by focussing on Bloch-Nordsieck inclusive cross sections, and using Weinberg's famous low energy theorem (for the gravitational case - for photons the result is actually due to Low :-))\cite{bnlw}, to determine the effect of adding additional soft gauge particles to an amplitude.  Although this completely resolves the IR problem for any real observation with a detector of finite energy resolution, it does not solve the mathematical problem of defining the Hilbert space on which the scattering operator acts. This question is central to any discussion of whether creation and evaporation of black holes is described by unitary quantum mechanics.

Progress on construction of the non-perturbative Hilbert space of $U(1)$ gauge theories and models of quantum gravity began with the work of Fadeev and Kulish\cite{fk} and continued since then in a slow trickle of papers\cite{trickle}.  Much of the recent focus of attention has been on the Bondi-Metzner-Sachs-van den Burg transformations\cite{bms}, which act on the Fock space of gravitons\cite{andyetal}.  In fact however, the Fadeev-Kulish construction and its gravitational analog proceed by constructing physical states that are invariant under BMSvB transformations and their $U(1)$ analog\cite{severporrati}\footnote{T.B. thanks Z. Komargodski for pointing out these papers and convincing him of their importance.}.  

Somewhat earlier, we made a proposal for the proper definition of the asymptotic Hilbert space of quantum gravity, in terms of algebras of current operators define on the past and future null cones $P^2 = 0$, which are the Fourier dual (w.r.t. the null coordinate on the boundary) of the conformal boundary of Minkowski space.  We suggested that the variable $P_{\mu}$ might be thought of as the spectrum of the BMSvB operators, but that they played no further role in the structure of the Hilbert space: all operators except the orbital part of Lorentz transformations commute with them.  The quantum information in the asymptotic Hilbert space is a representation of operators $Q_{\alpha}^i (P)$, which satisfy a generalization of the Awada-Gibbons-Shaw\cite{ags} super-BMS algebra.  The representations satisfy certain constraints that define ``exclusive Sterman-Weinberg jets"\cite{esw}, which imply in particular that there is non zero weight for the singular point $P = 0$ of the null cone.

The assertion is, that even when there are no IR divergences, there is a finite probability for gravitational scattering to produce states that contain ``infinite numbers of zero energy gravitons".  IR divergences occur when all the matrix elements of the non-perturbative scattering matrix in Fock space are zero. This occurs because in four dimensions the asymptotic trajectories of particles never have strictly zero acceleration, and accelerated particles radiate classical gravitational fields. That is, amplitudes for $n \rightarrow m$ with $n$ and $m$ finite are strictly zero in four dimensions.  However, even in higher dimension, where there are no IR divergences, the scattering operator is not unitary in Fock space. We will provide some evidence for this assertion below.

These considerations suggest the importance of amplitudes involving large numbers of low energy gravitons (and other massless particles), and one must revisit the question of whether AdS/CFT can reproduce these amplitudes in the limit $R/L_P \rightarrow\infty$.  We will argue that once
the number of gravitons, $n$, exceeds $R/L_P$ one begins to see deviations from flat space amplitudes.  Above a critical number of gravitons, whose dependence on $R/L_P$ varies with the number of dimensions, there is no way to get anything close to the flat space answer.  Indeed, above that critical value, the ``arena" gets swallowed by a stable AdS black hole. Between $R/L_P$ and the critical number, one always has finite probability for creating small evaporating black holes in the AdS calculation, while no such probability exists in the flat space amplitudes for most values of the kinematic invariants in flat space, as long as they are all $\ll$ Planck scale.

These observations also imply that the formation and evaporation of small black holes in AdS space, will not be able to reproduce the amplitudes for the qualitatively similar process in Minkowski space.  Note that we are not talking about gross properties like the temperature and entropy of the Hawking radiation, but rather about the distribution of quantum information in the asymptotic Hilbert space of quantum gravity.  It's obvious that an observer in the arena will have a hard time measuring anything to do with gravitons with momentum $\ll 1/R$, and indeed, flat space wavepackets formed only out of such momentum components will extend outside a region in flat space whose size is $R$.  Nonetheless, the question of the unitarity of the S matrix, including processes of black hole formation and evaporation, cannot be settled without getting these amplitudes right.

To put things another way, there is extreme non-uniformity in the way individual Fock space amplitudes, are approximated by AdS/CFT correlators, with the limit of a large number of legs not commuting with the $R/L_P \rightarrow\infty $ limit.  Inclusion of processes with an infinite number of legs is unavoidable in four dimensional Minkowski space, and we've argued, describes transitions into a finite probability subspace\footnote{By this phrase we mean a subspace of the Hilbert space such that the probability to make transitions into it starting from the Fock subspace is finite.}  of the Hilbert space in any space-time dimension. The question of unitarity of the flat space S matrix cannot be answered without properly treating transitions into the infinite particle part of the asymptotic Hilbert space, and we see no way of computing these transition amplitudes as limits of CFT correlators.

After establishing these results, we will turn to an argument that the infinite particle amplitudes are required to understand the ``unitarity cuts" in even the $ 2 \rightarrow 2$ amplitudes at low energy. We have put this last, because the discrepancy between AdS/CFT correlators and flat space amplitudes occurs for finite $n$ and finite momenta, at any finite $R$. To assess the relevance of that discrepancy to the question of non-perturbative unitarity we need results like this one about two to two amplitudes.

\section{The $2 \rightarrow n$ Amplitude}

The CHY\cite{CHY} formula for tree level $m$ graviton scattering in any dimension is
a remarkably compact packaging of the properties of perturbative quantum gravity. 
From these amplitudes, with $m = n + 2$ one can study the structure of $2 \rightarrow n$ amplitudes with c.m energy $E \ll M_P^{(D)}$ .   Here $D = 10$ or $11$ for most amplitudes we'd like to relate to AdS/CFT and $D = 6$ for the common $AdS_3$ models. The probability of having any pairwise impact parameters of order the Planck scale is very small.  If we divide 
the outgoing particles into two ``jets"\footnote{By this we mean only that we take two groups of particles that are most closely correlated in angle.  Most of these would not be considered jets by an experimentalist because the opening angles of the Sterman-Weinberg cones are very large.}, and compute the Mandelstam invariants we can make from the incoming graviton 
and outgoing jet momenta, the amplitude will fall as a power of the scattering angle $t/s$ and thus give a probability for small impact parameter $b$ that goes like $ (b\sqrt{s})^p$.  Nothing dramatic happens for impact parameters $b \sim L_P^{(D)}$  because the Schwarzschild radius of the c.m. energy is so much smaller than this.  Loop corrections to tree level amplitudes will not be important.  

Now contrast this with the situation in AdS space.  Following the prescription of \cite{polchsuss} we must act with smeared operators on the CFT vacuum, in such a way that we create a Feynman-Witten amplitude where $n + 2$ legs are all contained within the ``arena", a sphere of radius $< R$ surrounding some point in the bulk.  In $AdS_{d+1}/CFT$, the energy contained in the arena is bounded from below by $\frac{nd}{R}$ .  This means that there is a threshold for these processes at fixed $R$, which is definitely larger than $\sqrt{s} = 0$ and grows with $n$.  Furthermore, once $n \sim (R/L_P^{(D)})$ there is at least a Planck mass in the arena, and therefore a probability for black hole formation which is much larger than the flat space estimate above, simply because $\sqrt{s}$ is larger.  In other words, the convergence of multiparticle amplitudes to something resembling the flat space answers, is {\it non-uniform} in $R/L_P^{(D)}$ as this ratio goes to infinity.  

As $n$ increases above $R/L_P^{(D)}$ the discrepancy between flat space and AdS answers becomes more and more dramatic because we have a reasonable probability to create larger, meta-stable black holes.  The amplitudes will exhibit large time delays due to these resonances, and soft graviton radiation from the decay of relatively low temperature black holes, which is simply absent from the flat space answers.  Finally, as we increase $n$ further at fixed $R/L_P{(D)}$ we will get to the regime where there is a finite probability to create stable black holes. The first threshold, for small stable black holes\cite{gary}, which show up only in the microcanonical ensemble, occurs at $ E \sim M_P^{-1} (RM_P)^{\frac{(d - 1)(d - 3)}{2d - 3}}$.  This is followed by the creation of large stable black holes corresponding to typical states of the canonical ensemble of the boundary CFT, at temperatures above the Hawking-Page transition\cite{hp}.  In this ultimate situation, the ``arena" is completely swallowed by the black hole horizon.  

We can estimate the value of $n$ at which the final transition occurs, by CFT arguments.  The CFT lives on a $d - 1$ sphere of radius $R$.  The typical distance between operator insertions is thus $D(n) \sim R n^{ - \frac{1}{d - 1}}$ .  Tile the sphere with an $n$ independent set of spherical caps, whose centers are separated by solid angles of order $1$. The number of operators in each cap grows linearly with $n$.  In each cap we perform an OPE, to obtain a leading operator with a dimension of order $\Delta \sim n$ .  The dimension in each cap will be roughly the same, because the operators are distributed uniformly.   Once $\Delta$ is in the range where the asymptotic entropy formula
\begin{equation} S(E) = c (RE)^{\frac{d - 1}{d}} , \end{equation} with \begin{equation} c \sim \bigl{(}\frac{R}{L_P^{(d + 1)}}\bigr{)}^{\frac{d - 1}{d}} , \end{equation} is valid, then the degeneracy of operators with dimension $\Delta$ is that of a large black hole. Note that if the distribution of operators is non-uniform then we will reach this asymptotic regime in {\it some} caps for smaller values of $n$. The range of validity can be calculated from the entropy formula by insisting that the temperature be higher than the Hawking Page value.  This gives
\begin{equation} n > \bigl{(}\frac{R}{L_P^{(d + 1)}}\bigr{)}^{d - 1} . \end{equation}   The same estimate comes from bulk physics by asking when the Schwarzschild radius of the c.m. energy necessary to create the $n$ particle state, equals the size of the arena.

We emphasize that there will be very clear discrepancies between the AdS and flat space calculations at much lower values of $n$.  In weakly coupled string regimes, the situation is even worse because production of string states will set in for values of $n$ parametrically smaller than this estimate, by powers of the string coupling.  No such production occurs for $2 \rightarrow n$ processes with any value of $n$, at low c.m. energy in flat space.

\section{Zero Energy Singularities in $2 \rightarrow 2$ Scattering}

Let us consider a model of quantum gravity in $D$ Minkowski dimensions, where $D$ is sufficiently large that there are no IR divergences.  As an example, we could consider M-theory, the quantum completion of the model whose low energy amplitudes are well approximated by the tree diagrams of the 11 dimensional supergravity Lagrangian.  Bandos\cite{bandos} has recently written an elegant set of recursion relations for the general n-point tree amplitude in this model.  It contains a subsector of pure graviton scattering, identical to the CHY formula we mentioned above.   We will first discuss models which have no small dimensionless expansion parameter.

The exact transition operator $T$ of this model satisfies the unitarity equation.
\begin{equation} i(T - T^{\dagger}) = T T^{\dagger} . \end{equation} Take the matrix element of this equation between states containing two gravitons, with momenta and polarizations $k_a , \epsilon_a$, where all momenta are much smaller than the Planck scale, $k_a / M_P \ll 1$.  Insert a complete set of intermediate states and write 
\begin{equation} {\rm Im}\ T (k_a , \epsilon_a) = \sum_s \langle k_3 , \epsilon_3 , k_4, \epsilon_4 | T | s \rangle \langle s | T^{\dagger} |  k_1 , \epsilon_1 , k_2, \epsilon_2 \rangle , \end{equation} where $| s \rangle$ are a complete set of intermediate states.   Energy momentum conservation restricts these states to have very low momenta, so that we can calculate the matrix elements $ \langle k_i , \epsilon_i , k_j, \epsilon_j | T | s \rangle $ using the tree approximation, up to terms which go to zero like powers of the Mandelstam invariants $s$ and $t$.  
The intermediate states $| s \rangle$ can contain particles other than gravitons.  However, if we take the forward direction, each intermediate state contributes positively to this sum.  Near the forward direction there cannot possibly be complete cancelation, and analyticity of the amplitude away from $t = 0$ guarantees that there will not be a cancellation of different contributions, except at isolated values of $t$.  We can therefore obtain a bound on the imaginary part of the amplitude by using the CHY formula, in a quite general model.  In M-theory we could in principle use the recursion relations of Bandos to include all contributions to leading order in $s$ and $t$, and obtain a tighter lower bound on the amplitudes of $n$-particle intermediate states.  

In either case, we can see two very general features of the contribution from intermediate states containing $n$ identical particles:

\begin{itemize}

\item The $n$ particle amplitude has a complicated branch point at $s = 0$, different for every $n$.

\item There is a factor of $n!$ in the $n$ particle contribution to the amplitude, which comes from permuting the internal legs. However, although the full Feynman diagram which has the $n-$particle cut does in fact behave like $n!$, the phase space integral over the cut has a compensating factor of $1/n!$\footnote{We thank S. Raju for pointing this out to us.}. We'll explore the consequences of this below.

\end{itemize}

These two features imply that the two to two amplitude has a complicated singularity at $ s = 0$.  While we'd originally argued that the low energy unitarity argument alone implied a divergence of the perturbation series, the $1/n!$ phase space suppression shows only that the series has a finite radius of convergence.  Positivity of the coefficients near the forward direction implies that the singularity is on the positive real $s$ axis, signaling the production of meta-stable states.  In perturbative string theory these are fundamental strings, D-branes and eventually black holes.  In a more generic region of moduli space, it would only be black hole states.  

The real question of what sort of soft graviton states are produced in gravitational scattering thus involves complicated questions of the classical radiation emitted in multi-black hole systems, the re-scattering of the high energy gravitons emitted in the late stages of Hawking decay {\it etc.} .  In finite orders of string perturbation theory of course we only see Fock space states, but that perturbation series is not even Borel summable.

String perturbation theory is enough to reveal the complicated nature of the question of soft graviton final states.  Consider scattering of two gravitons at energies well above the string scale, but below the Planck scale.  One can compute the amplitudes to produce the exponentially large number of perturbative string states that are available at those energies, and they are all finite.  Each of these states will decay, typically by emitting low energy, short strings, which include gravitons, and cascading down to the ground state, leading to a plethora of soft gravitons.  There will typically be interactions between the decay products, producing even more gravitons.

Consider for example the production of two long semi-classical strings.  These will have long range gravitational interactions, causing them to orbit around each other, producing gravitational radiation in the process.  For some initial configurations the orbits might be bound states.  The strings will spiral around each other and collide again to produce further excitations, which will themselves decay, {\it etc.}.  Thought of as diagrams in string perturbation theory, these processes, described classically at leading order, involve arbitrarily many powers of the string coupling.  As noted above, string perturbation theory does not converge and there is no principled way to claim that the non-perturbative final state lies in Fock space.  The failure of any amplitudes to represent Fock space matrix elements in four dimensions, suggests strongly that the existence of transitions out of Fock space is present in any dimension, and we here see explicit mechanisms by which that might happen. Away from the weak coupling region of moduli space, the thresholds indicated by our unitarity argument have to do with black hole production, and we can repeat the above arguments with black holes replacing macroscopic strings.  

Consequently it seems highly implausible that the low energy expansion of the two to two amplitude is convergent.  Furthermore, the singularity at $s = 0$ cannot be explained by any intermediate state in Fock space.
Any such state would have a convergent expansion in terms of states containing a finite number of particles.  It could not have the singularity structure of the true amplitude, and it could not explain the divergence of the low energy expansion.  This proves that, even in high dimension and with maximal SUSY, models of quantum gravity require a re-definition of the asymptotic Hilbert space.  The correct non-perturbative Hilbert space must allow for states that we can describe, by abuse of language, as containing ``infinite numbers of zero energy gravitons".  In previous publications\cite{previous} we have proposed a definition of the Hilbert space as a representation of an algebra of fermionic currents $Q_{\alpha}^i (P), Q_{\alpha}^i (\tilde{P}) $, where $P$ lies on the null cone $P^2 = 0$, $\tilde{P}$ is the parity reflection of $P$, and the scattering operator maps the algebra with $P_0 < 0$ into that with $P_0 > 0$. The fermionic operators are half measures on the cone, with non-zero support at $P = 0$.  We believe that it is obvious, from our discussion in the previous section, that CFT correlators cannot contain these singularity structures at {\it any} finite value of $R$.  Therefore, {\it the non-perturbative scattering operator of quantum gravity in Minkowski space cannot be obtained as a limit of CFT correlation functions}.

Sophisticated readers will wonder why the same sort of arguments do not apply to models with exact Nambu-Goldstone bosons of a non-abelian group. We will address this question in the conclusions.  

We close this section with a comment about models which {\it do} have a dimensionless expansion parameter, including all versions of perturbative string theory.  At any finite order in the $g_S$ expansion, we will see none of the structure revealed by the argument above.  It is well known that the $g_S$ expansion is asymptotic and not Borel summable\cite{shenkergrossmende}.  The lesson from our argument is that the ``summation" of the $g_S$ expansion will require a qualitative change in the description of the non-perturbative Hilbert space.

In \cite{previous} we described the Hilbert space as a space of ``exclusive Sterman-Weinberg jets" , where each amplitude was parametrized by the representation of the $Q_{\alpha} (P)$ operators, with $P \neq 0$ on a collection of finite spherical caps in the sphere at infinity, as well as the state of the $ P = 0$ operators in the complements of those caps.  We believe that the angular openings of those caps go to zero as $g_S \rightarrow 0$, and that the zero momentum variables decouple, but it is not clear to us how to reconstruct the correct S matrix from string perturbation theory.  

\section{Conclusions}

The previous arguments suggest that the non-perturbative definition of the Scattering operator in models of quantum gravity is not an S-matrix in Fock space.  They also demonstrate that AdS/CFT will probably not help to elucidate the true structure of this operator.  Matrix Theory\cite{bfss} remains a non-perturbative tool, which might help us to formulate the proper framework.  The issue of zero energy gravitons takes on an interesting form in Matrix Theory.  The S-matrix in that formulation is a Scattering operator on the moduli space of a supersymmetric quantum mechanical system, with gauge group $U(N)$.  The moduli space is the Coulomb branch of the gauge theory, broken down
to $U(n_1) \times \ldots U(n_k)$.  $N$ is viewed as the fixed total longitudinal momentum in a Discrete Light Front Quantization.   The Matrix Theory conjecture is that by taking the $N\rightarrow\infty$ limit, and restricting to the subspace of the Hilbert space with energy scaling like $1/N$ in that limit, one retrieves the Lorentz invariant Scattering operator on Minkowski space of varying dimensions, with maximal and half maximal SUSY.  

At finite $N$ the Matrix Theory scattering operator is an operator on an asymptotic Hilbert space containing at most $N$ supergravitons.  As $N \rightarrow \infty$ sub-blocks of the matrix variables with $n_i = k N$, $ 0 < k < 1$, behave like finite energy supergravitons with longitudinal fraction $k$.
However, there are also states with energy $\sim 1/N$ with {\it transverse momentum} $| {\bf P}_T | \leq \frac{1}{\sqrt{N}} $ and block sizes $n_j$ which remain finite as $N \rightarrow \infty$.  These are the zero momentum super-gravitons in the limit, and the limiting S operator will operate on a space containing arbitrary amounts of quantum information describing the states of these super-gravitons at various angles.  Thus, Matrix Theory, for those models to which it applies, may provide a systematic non-perturbative treatment of gravitational scattering in Minkowski space.

We have remarked elsewhere\cite{previous} that the Holographic Space-time formalism (HST) was a natural restriction of the current algebra of the $Q_{\alpha} (P)$ to finite causal diamonds, and provides a more general finite approximation to gravitational scattering.  In this context, it is interesting to recall our recent investigation into the relation between HST and the AdS/CFT correspondence\cite{hstads}.  We found that, for CFTs with large radius CFT duals and living in the regime where $g_S \sim 1$, the physics in the ``arena" could not be extracted from the dynamics described by the CFT Hamiltonian.  That result is entirely consistent with the results of this paper.

Finally, let us address the question of the proper venue for the physics of Nambu-Goldstone bosons, since our S-matrix arguments about zero momentum singularities apply equally well to that case.  The major difference is that theories of Goldstone bosons are quantum field theories, and have perfectly well defined Green's functions at finite points\cite{elitzurdavid}.  Thus one expects that a version of the Wightman/Osterwalder-Schrader reconstruction theorem would allow one to construct the Hilbert space.  Strictly speaking, one might first have to break the symmetry explicitly, to give a theory with the mass gap required by those theorems, and then take the limit, but we know that the limiting Green's function exist.  The question of whether there's a scattering operator acting in Fock space is somewhat different.  Questions related to this were studied cursorily in the ancient literature\cite{weinbergrosensweig}, but to our knowledge the problem has never been completely solved.

In passing we note that a similar strategy does not seem to work for $U(1)$ gauge bosons.  We can introduce a Higgs field to spontaneously break the symmetry, but if we try to deform the Higgs potential to go to the Coulomb phase, we encounter the Coleman-Weinberg, Halperin-Lubensky-Ma, fluctuation induced first order phase transition.  Before the Higgs phase disappears, it becomes a meta-stable excitation of the symmetric ground state.   The scattering operator computed in the Higgs phase will cease to be unitary in the Fock space of massive photons and Higgs bosons, because of transitions in which the meta-stable Higgs vacuum decays to the symmetric ground state.

\vskip.3in
\begin{center}
{\bf Acknowledgments }\\
TB thanks Z. Komargodski for discussions clarifying the role of the BMS algebra in the definition of the non-perturbative Hilbert space of Quantum Gravity. The work of T.Banks is {\bf\it NOT} supported by the Department of Energy. The work of W.Fischler is supported by the National Science Foundation under Grant Number PHY-1620610. A preliminary version of this work was presented at the NatiFest conference at the Institute for Advanced Study in September 2016.  Video may be found at the IAS website. TB thanks the organizers of that conference, especially J. Maldacena, for the opportunity to speak there. 
\end{center}

\end{document}